# Synthesis and Properties of Bismuth Ferrite Multiferroic Nanoflowers

K. Chybczyńska[a], P. Ławniczak[a], B. Hilczer[a], B. Łęska[b], R. Pankiewicz[b], A. Pietraszko[c], L. Kępiński[c], T. Kałuski[d], P. Cieluch[d], F. Matelski[e] and B. Andrzejewski[a*]

[a] Institute of Molecular Physics
Polish Academy of Sciences
Smoluchowskiego 17, PL-60179 Poznań, Poland
* corresponding author: bartlomiej.andrzejewski@ifmpan.poznan.pl

[b] Faculty of Chemistry
Adam Mickiewicz University
Umultowska 89b, PL-61614 Poznań, Poland

[c] Institute of Low Temperature and Structure Research
Polish Academy of Sciences
Okólna 2, PL-50422 Wrocław, Poland

[d] Research Centre of Quarantine, Invasive and Genetically Modified Organisms
Institute of Plant Protection – National Research Institute
Węgorka 20, PL-60318 Poznań, Poland

[e] Faculty of Technical Physics
Poznan University of Technology
Nieszawska 13a, PL-60965 Poznań, Poland

*Abstract*— **The method of microwave assisted hydrothermal synthesis of bismuth ferrite multiferroic nanoflowers, their mechanism of growth, magnetic as well as dielectric properties are presented. The nanoflowers are composed of numerous petals formed by $BiFeO_3$ (BFO) nanocrystals and some amount of amorphous phase. The growth of the nanoflowers begins from the central part of calyx composed of only few petals towards which subsequent petals are successively attached. The nanoflowers exhibit enhanced magnetization due to size effect and lack of spin compensation in the spin cycloid. The dielectric properties of the nanoflowers are influenced by water confined to crystal nanovoids resulting in a broad dielectric permittivity maximum at 200 K ÷ 300 K and also by Polomska transition above the temperature of 450 K.**

## I.  INTRODUCTION

Recent development of nanoscience and nanotechnology has brought about the discovery of miscellaneous forms of nanoobjects. Among them one can mention nanopowders, nanotubes, nanowires, nanorods, nanosheets, nanoclusters, nanocones and even such sophisticated nanoobject like nonobowlings, nanobottles and nanonails. One of the most rarely met nanobjects are nanoclusters in the form of nanoflowers reported in a nice review by Kharisov [1]. Nanoflowers can be composed of various elements and compounds like: metals (Au [2], Ni [3], Zn [4], Sn [5], Co[6]) and carbon [7, 8], metal oxides (ZnO [9, 10], MgO [11], CuO [12], $\alpha$-$MnO_2$ [13], $SnO_2$ [14]), hydroxides and oxosalts ($Mg(OH)_2$ [15], $Cu_2(OH)_3Cl$ [16]), sulphides, selenides and tellurides (CdS [17], ZnSe [18], PbTe [19]), nitrides and phosphides (lnN [20], GaP [21]), organic and coordination compounds ($CdQ_2$ complexes [22], alkali earth phenylphosphonates [23]).

The nanoflowers exhibit different dimensions from a few dozen of nanometers (about 40 nm for $In_2O_3$ nanoflowers [24]) up to micrometers (about 50 μm for $SnO_2$ nanoflowers [14]) and also a variety of morphological details. For example, perforation of petals in NiO [25] nanoflowers, sheet-like petals [2] in Au nanoflowers, hollow cores [7] in carbon nanoflowers, plate-like and brush-like shapes composed of pyramidal nanorods in ZnO nanoflowers [9] have been observed. Other examples are MgO nanoflowers consisting of nanofibers [11], CuO nanoflowers with petals branched into tips [12], $\alpha$-$MnO_2$ with nanocrystalline petals [13], nanobelt-like petals in $MoO_3$ [26], hexapetalous snowflake-like $Cu_7S_4$ nanoflowers [27], downy-velvet-flower-like nanostructures of PbS [28], vaselike $CuInS_2$ nanostructures [29], hexangular shapes of InN compound [30] and at last bundles of nanorods of $CdQ_2$ complexes [22].

Methods of synthesis of nanoflowers include reduction of metal salts [2, 3], electrodeposition [31], catalytic pyrolysis method [32], zero-valent metal oxidation or decomposition of compounds to obtain metal oxides [10], chemical vapor transport and condensation [14, 30, 33], calcination [25] and many others. However, it seems that the solvothermal and hydrothermal techniques are here especially effective [5, 13, 15, 27, 28, 34-37]. In some cases the hydrothermal process was activated by microwave heating [3, 17, 38].

The nanoobjects in form of nanoflowers are important from the point of view of future applications because they exhibit excellent electrocatalytic activity [39], high-dielectric permittivity, [11], photocatalitic activity [34] and they can be used as excellent field emitters [40, 41], effective solar cells [15] or amperometric biosensors [42].

In this work, we report for the first time on the microwave assisted hydrothermal synthesis, structure, electric and magnetic properties of bismuth ferrite $BiFeO_3$ (BFO) nanoflowers. Bismuth ferrite, even in bulk is an unusual material because it belongs to magnetoelectric (ME) multiferroics that exhibit simultaneously charge and magnetic ordering with some mutual coupling between them at room temperature [43-46]. These materials have recently attracted world-wide attention because of their interesting physical properties and large technological potential to be applied as four-state memory and in spintronic devices. Other important applications of BFO in piezoelectric devices, as THz radiation emitters or catalysts are not related to ME effects [47].

From the point of view of structure BFO oxide is a rhombohedrally distorted perovskite with space group $R3c$ at room temperature. The ferroelectric (FE) properties appear in BFO below ferroelectric Curie temperature $T_C$=1100 K, whereas the antimagnetic ordering together with weak ferromagnetic (FM) moment appear below the Néel temperature $T_N$=643 K. These FE and FM properties result from the charge ordering caused by lone electron pairs of $Bi^{3+}$ ions and from the complex ordering of $Fe^{3+}$ spins, respectively. BFO compound exhibits an antiferromagnetic (AFM) G-type ordering and superimposed long range incommensurate cycloidal modulation with the period $\lambda$=62 nm [45, 48]. The spin cycloid can propagate along three equivalent crystallographic directions [1,-1,0], [1,0,-1] and [0,-1,1] (pseudocubic notation). The spins in the cycloid rotate in the plane determined by the direction of the vector of cycloid propagation and the [1,1,1] direction of the vector of spontaneous electric polarization. Weak FM moment in this AFM compound originates from an interaction similar to Dzyaloshinskii-Moriya one which forces small canting of the spins out of the rotation plane. The weak FM moment increases due to the size effect apparent when the dimensions of BFO particle become comparable to the modulation period $\lambda$. In this case, the lack of magnetic moment compensation in the spin cycloid is responsible for enhanced magnetic properties of BFO nanoobjects, which is much desired from the point of view of ME coupling magnitude [49].

The aim of this work is to present the method of microwave assisted hydrothermal synthesis of BFO multiferroic nanoflowers as well as to study their basic ferroelectric and magnetic properties with a special attention to the size effect.

## II. EXPERIMENTAL

### A. Sample Synthesis

BFO ceramic powder-like samples were synthesized by means of microwave assisted hydrothermal Pechini method [50, 51] using bismuth and iron nitrates as precursors: $Bi(NO_3)_3 \cdot 5H_2O$, $Fe(NO_3)_3 \cdot 9H_2O$. The nitrates together with $Na_2CO_3$ were added into a KOH solution of a molar concentration of 6 M. The mixture was next transferred into a Teflon reactor (XP 1500, CEM Corp.) and loaded into a microwave oven (MARS 5, CEM Corp.). The reaction was carried out at the same temperature for all samples (200 $^0$C) for a short time equal to 20 min (samples labeled ST) or for one hour which was the long time synthesis (samples labeled LT). After processing, BFO powders were first cooled to 20 $^0$C, next collected by filtration kit, rinsed with distilled water and placed in a dryer for 2 h. The final products were brown powders of BFO agglomerates. In this way a set of ST and LT samples composed of grains in the shape of nanoflowers was synthesized. Another sample, labeled STC was obtained after air calcination of ST sample at 500$^0$C for 1h.

### B. Sample Characterization

The crystallographic structures and phase compositions of the BFO samples were studied by means of x-ray diffraction method (XRD) using an ISO DEBYEYE FLEX 3000 diffractometer equipped with a HZG4 goniometer in Bragg-Brentano geometry and with a Co lamp ($\lambda$=0.17928 nm). The morphology and microstructure of the samples were studied by FEI NovaNanoSEM 230 scanning electron microscope (SEM) and also by Philips CM20 SuperTwin transmission electron microscope (TEM). Mmagnetometric measurements were performed using a Vibrating Sample Magnetometer (VSM) probe installed on the Quantum Design Physical Property Measurement System (PPMS) fitted with a superconducting 9T magnet. Dielectric properties and electric conductivity of the BFO samples were studied by means of an Alpha-A High Performance Frequency Analyzer (Novocontrol GmbH) combined with a Quatro Cryosystem for the low temperature control. Before the measurement, the BFO fine powders were pressed to a compact form of pellets of dimensions 5.15 mm in diameter and ca. 1 mm in thickness, onto which silver paste electrodes were deposited. The measurements were recorded on heating at the rate of 0.25 K/min while the frequency was varied from 1 Hz to 1 MHz at the oscillation voltage of 1 V.

## III. RESULTS AND DISCUSSION

The time of microwave reaction and the concentration of KOH were found to be crucial factors determining the shape of the BFO agglomerates obtained. At the KOH concentration of 6 M and after 20 min of the reaction, numerous agglomerates in the shape of nanoflowers were formed, (ST sample).

The processing time of about 1 h allowed getting large but singular BFO nanoflowers (LT sample). Besides the nanoflowers, other forms like large BFO microspheres were present. Higher molar concentration of KOH equal to 10 M and the synthesis time of 30 min promoted formation of microcubes or spheres instead of the nanoflowers.

Figs. 1 and 2 present SEM micrographs of ST sample obtained at the KOH concentration of 6 M and 20 min synthesis. This sample contains only agglomerates in the form of nanoflowers, which are indeed very similar to numerous flower-like structures reported in ref. [1]. The mean size of the nanoflowers is of about 15 μm. The nanoflowers are not perfectly spherical and exhibit kind of hollows where the petals are packed less dense. The structure of the hollows and details of petals ordering in their vicinity is shown in Fig. 3 which presents a small nanoflower very similar to a rose. The petals in this nanoflower have more irregular shapes than in the large nanoflowers and form a kind of hole, or using the botanic terminology "bottom of calyx" near the center. The size of this hole is less than 1 μm. The bottom of calyx is very pronounced in the structure shown in Fig. 4 which presents the BFO nanoflower "in statu nascendi". The structure of this nanoflower indicates the process of its formation. One can assume that at early stages of growing a central part composed of only a few petals (or even four as in Fig. 4) is formed. The petals are connected at the edges and form a kind of rectangular cage. Next, around this structure, successive petals nucleate. For small flowers, like in Fig. 4 the petals are mainly located in the plane perpendicular to the flower axis and perpendicularly to the petals in the center. In the bigger flowers, the petals can take various orientations. We assume that the big, spherical nanoflowers in Figs. 1 and 2 also contain cages inside but they are well masked by numerous petals around them. This explanation is supported by the observation that the holes in the center of the nanoflower are apparent mainly for small nanoflowers (about 5 μm in diameter in Fig. 3 or about 2.5 in diameter μm in Fig. 4) being in the initial stages of formation. The big nanoflowers are always almost spherical.

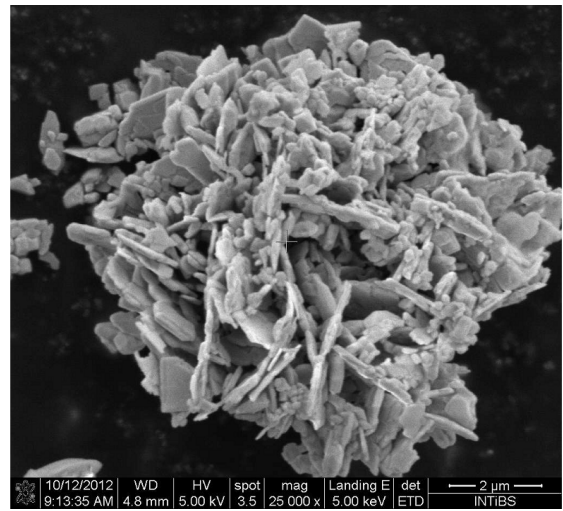

Figure 2. The SEM micrograph presenting the order of petals near the hollow part of one of the selected small nanoflowers in ST sample.

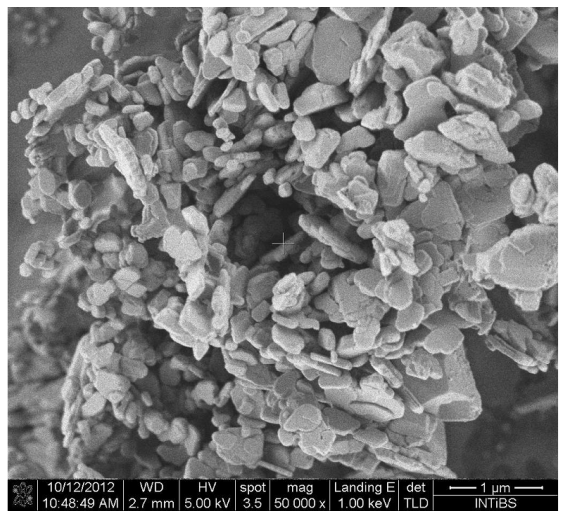

Figure 3. SEM micrograph of the nanorose found in the ST sample.

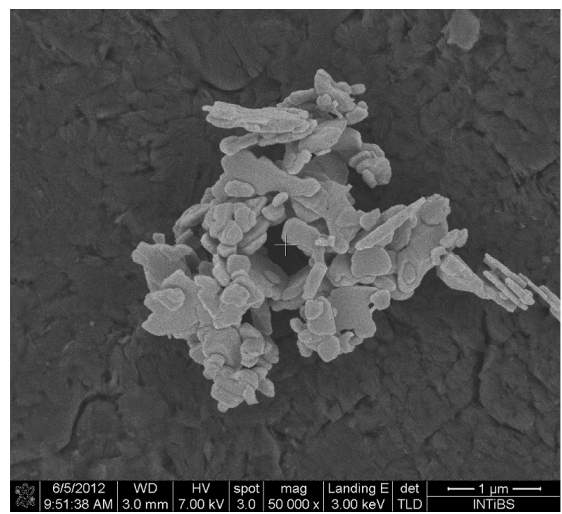

Figure 4. SEM micrograph of the initial stage of formation of a nanorose observed in the ST sample. Four petals in the center of the nanorose form a cage.

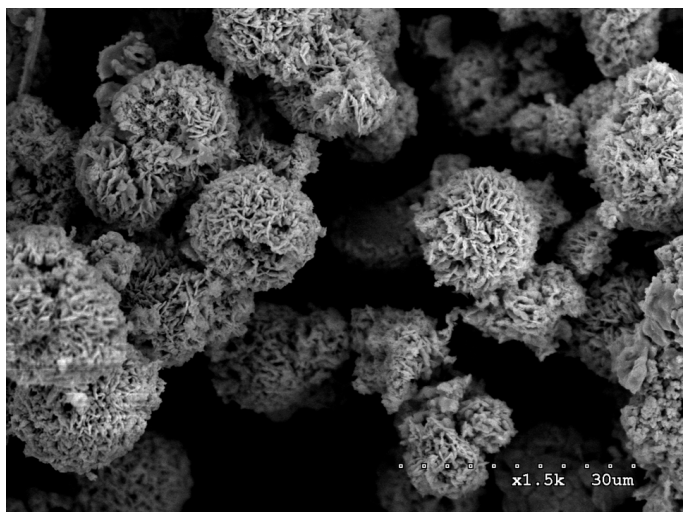

Figure 1. The SEM micrograph of ST sample composed of nanoflowers only.

Fig. 5 presents SEM micrograph of LT sample processed for 1 h. The image shows large, well formed spherical nanoflowers of about 20 μm in diameter. The nanoflowers are composed of hundreds of closely packed petals. Besides nanoflowers, in this sample there are numerous other structures like microspheres and other irregular forms. It was observed that the long time of synthesis and crystallization promoted formation of large nanoflowers and also more dense structures composed of massive BFO microcrystals. Probably, the spherical nanoflowers transform to microspheres because of the crystal growth of petals as the reaction proceeds.

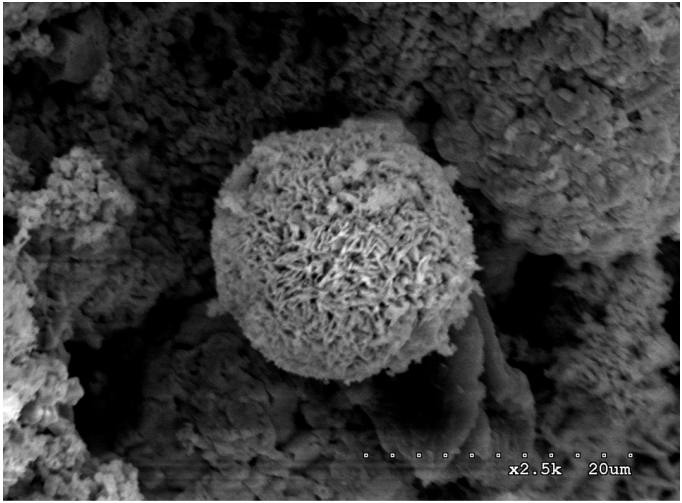

Figure 5. The SEM micrograph of a BFO nanoflower found in LT sample.

The thickness of petals of BFO nanoflowers varies from about 30 nm to 200 nm as presented in the histogram in Fig. 6. Most of the petals have thickness in the range 50÷100 nm. The distribution of the petals thickness can be fitted using log-normal function well describing granular random systems [52]:

$$f(D) = \frac{1}{\sqrt{2\pi\sigma^2}} \frac{1}{D} \exp\left(-\frac{1}{2\sigma^2} \ln^2\left(\frac{D}{\langle D \rangle}\right)\right) \qquad (1)$$

where $\langle D \rangle$ denotes the median thickness of the petal and $\sigma$ is the distribution width. The best fit of eq. (1) to the data presented in the histogram is obtained for $\langle D \rangle$=78(3) nm and $\sigma$=0.40(3). The maximum value of the distribution $f(D)$ of petal thickness appears at 67 nm.

Nanostructure of the petals of nanoflowers was analyzed by means of TEM and the image is presented in Fig. 7. It turns out that the petals are composed of BFO blocks with dimensions exceeding 100 nm. Some of them, like the biggest block in Fig. 7, exhibit well crystallized bismuth ferrite phase, which was confirmed using selected area diffraction (SAD) method (see Fig. 8). An example of the agglomerate containing very small BFO nanocrystals probably embedded in the amorphous matrix is shown in Fig. 9. Wide diffraction rings in the SAD pattern of this agglomerate presented in Fig. 10 confirm that the size of the nanocrystals does not exceeds a few nm. Agglomerates composed of bigger BFO crystallites with dimensions about 10-20 nm, embedded in the amorphous phase are also observed.

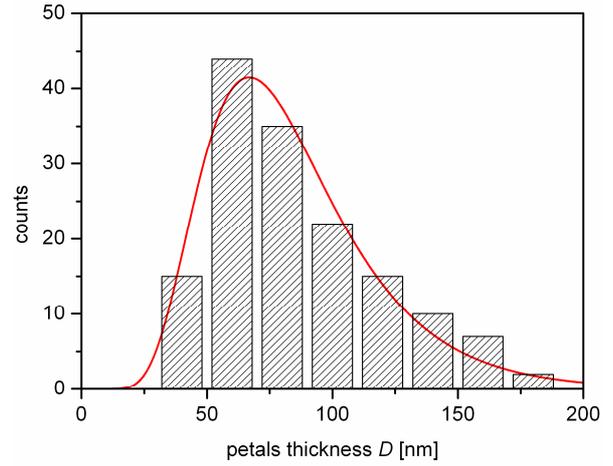

Figure 6. The histogram of thickness of petals of BFO nanoflowers. The solid line is the best fit of the log-normal distribution eq. (1) to the data.

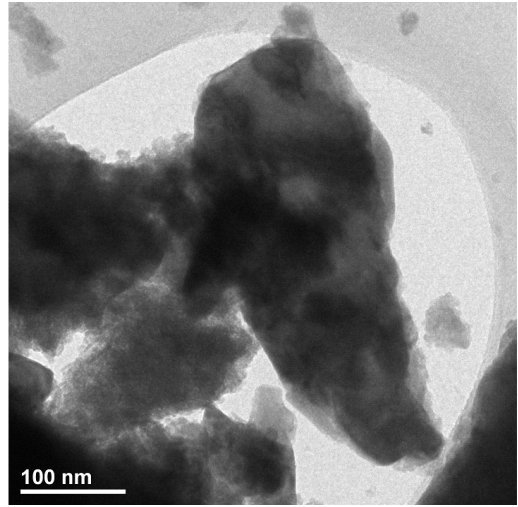

Figure 7. The TEM micrograph of the BFO crystallites from the ST sample.

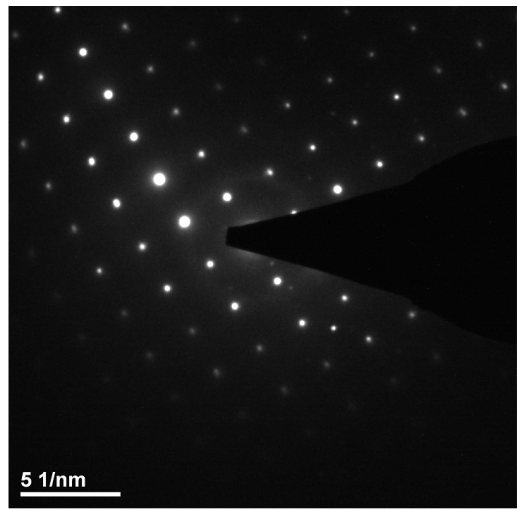

Figure 8. The SAD of the biggest crystallite in Fig. 7.

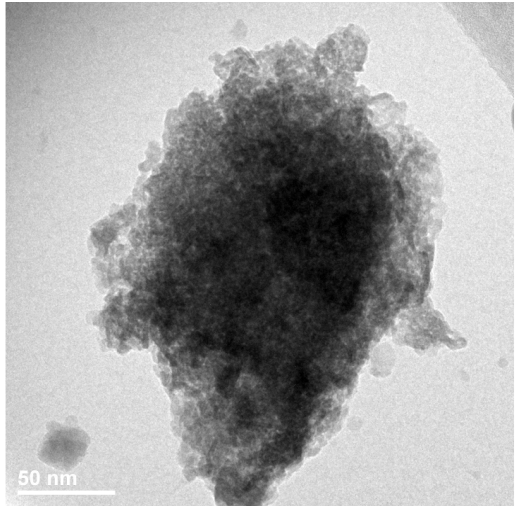

Figure 9. The TEM micrograph of the BFO agglomerate in the ST sample.

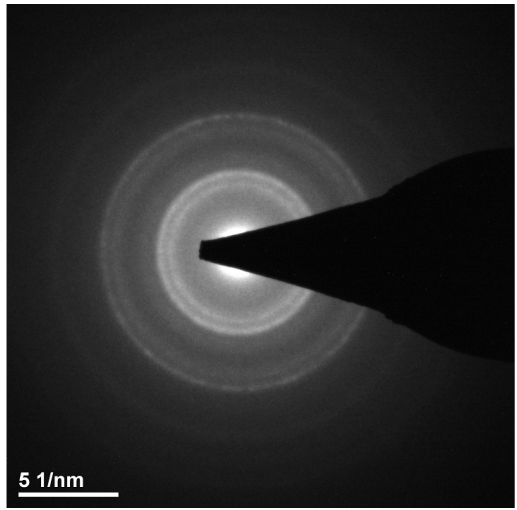

Figure 10. The SAD pattern of the BFO agglomerate presented in Fig. 9.

The crystalographic structure of the BFO samples was studied by means of x-ray diffraction at room temperature. The XRD patterns for as-prepared powders containing nanoflowers synthesized during 20 min (ST sample) are presented in panel "a" of Fig. 11, for the sample synthesized during 1h (LT sample) in panel "b" and for the calcined ST sample (STC sample) in panel "c". The solid lines in Fig. 11 correspond to the best fits by means of Rietveld method performed using FULLPROF software to the experimental data represented by open points. The lines below the XRD data illustrate the difference between the data and the fit. The vertical sections indicate the positions of individual Bragg peaks. Analysis of these XRD patterns reveals the presence of the BFO rhombohedral phase with *R3c* space group in all samples. The parameters of the hexagonal crystallographic cell in ST sample are: $a=b=5.577(1)$ Å and $c=13.862(2)$ Å. The unit cell parameters of STC sample obtained after air calcination of ST sample at 500 $^0$C for 1 h are unchanged, because: $a=b=5,576(1)$ Å and $c=13,858(2)$ Å. However, the LT nanoflowers obtained in the long-time synthesis exhibit smaller crystallographic cell as compared to those in the ST and STC samples: $a=b=5.572(2)$ Å and $c=13,847(5)$ Å. Probably, these parameters can be related to irregular BFO agglomerates present in this sample. The peak at about 34 deg marked by asterisk in Fig. 11 originates from a small content of $Bi_{25}FeO_{40}$ parasitic phase. The crystallographic parameters of BFO cubes and spheres are close to those of the nanoflowers present in ST and STC samples.

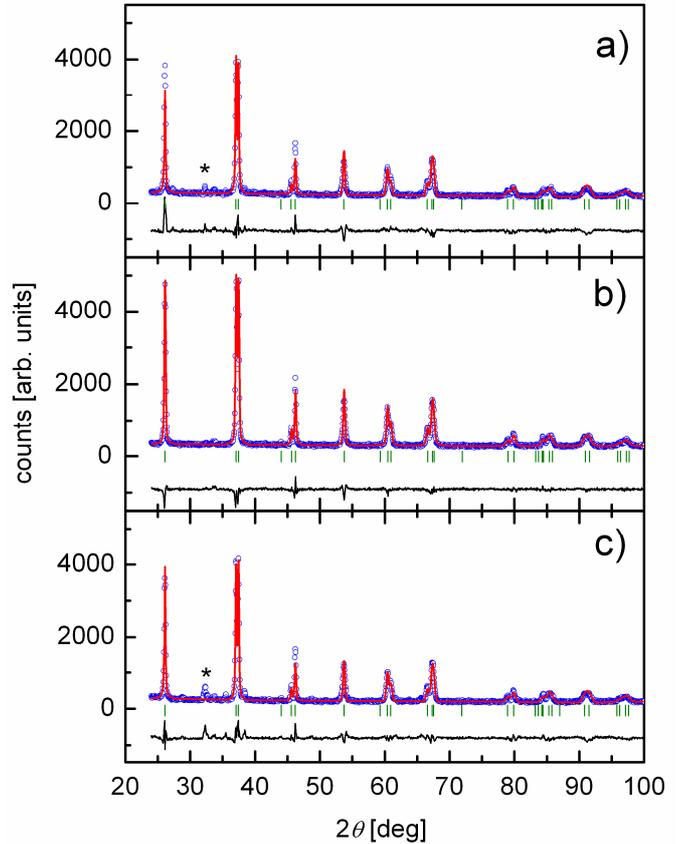

Figure 11. The XRD-pattern of ST sample prepared during 20 min processing (panel a), LT sample synthetized in 1 h (panel b) and calcined STC sample (panel c). The solid line is the best fit to the experimental data represented by open points. The line below the fits represent the difference between the data and the fit. The vertical sections represent positions of the Bragg peaks. The asterisk indicates the trace of $Bi_{25}FeO_{40}$ phase.

Mean size of crystallites in the nanoflowers was estimated using the Scherrer's equation [53]: $D=K\lambda/\beta\cos\Theta$, where $D$ is the crystallite size (in nm), $\beta$ is the half-width of the diffraction peak and $\Theta$ represents the position of the Bragg peak. The constant K in the Scherrer's equation depends on the morphology of the crystal [54] and here it was assumed that K=0.93 as in the Scherrer report [53]. Mean size of the crystallites in ST and LT samples was assessed to amount to about $D=37$ nm. It seems however, that the calcination slightly increased the mean size of crystallites in STC sample to about $D=39$ nm. The growth of crystallites is probably due to crystallization of the amorphous phase. The crystallographic parameters of the nanoflowers studied are collected in Table I.

**Table I. Hexagonal unit-cell parameters and the crystallite size of BFO nanoflowers**

| sample | $a=b$ [Å] | $c$ [Å] | $\alpha, \beta, \gamma$ [deg] | $D$ [nm] |
|---|---|---|---|---|
| LT nanoflowers | 5,572(2) | 13,847(5) | $\alpha=\beta, \gamma=120^0$ | 37 |
| ST nanoflowers | 5,577(1) | 13,862(2) | $\alpha=\beta, \gamma=120^0$ | 37 |
| STC nanoflowers | 5,576(1) | 13,858(2) | $\alpha=\beta, \gamma=120^0$ | 39 |

The mean size of the crystallites can be compared to the thickness of the nanoflower petals whose thickness distribution is presented in the histogram Fig. 6. It is evident that the mean size of the crystallites well corresponds to the thickness of the finest petals, equal to about 40 nm. Therefore, one can assume that the thinnest petals are composed of a single layer of BFO crystallites.

Besides the morphology and the structure of BFO nanoflowers we have also studied their magnetic and dielectric properties i.e. basic parameters characterizing the multiferroics. Fig. 12 presents the magnetic hysteresis measurements at 300 K for the samples composed of ST, STC and LT nanoflowers.

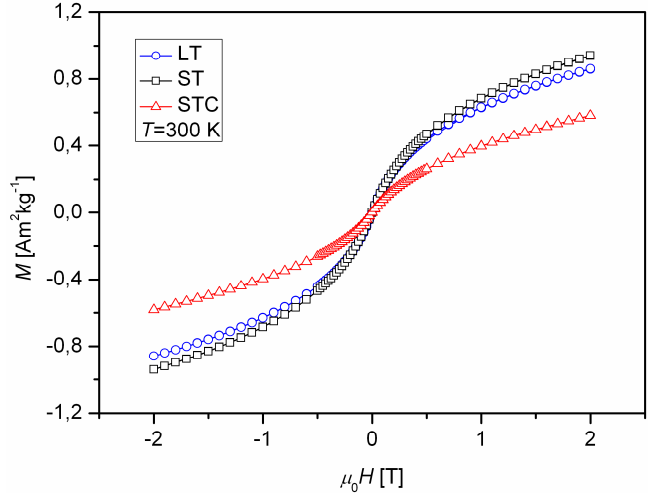

Figure 12. Magnetic hysteresis curves $M(H)$ recorded at 300 K for the BFO nanofloweres.

The shape of the magnetization curves $M(H)$ is similar for all samples however the differences in the value of magnetization are substantial. The highest magnetization was measured for the samples containing LT and ST nanoflowers and it was almost 3 times higher than the BFO bulk magnetization of about 0.3 Am$^2$/kg reported by Park [49] (all measurements performed at 2 T). The enhanced magnetization of nanoflowers can be related to the size effect because of the uncompensated spins near the surface [49]. This effect appears when the size of the BFO particle is lower than the period of the spin cycloid $\lambda$=62 nm. Indeed, the thickness of the finest petals evaluated for 40 nm is smaller than $\lambda$ length and the nanoflowers exhibit higher magnetization equal to 0.94 Am$^2$/kg for ST and to 0.86 Am$^2$/kg for LT samples (both measured at 2 T) than that of the bulk BFO material. This result well corresponds to the data published in earlier report [49] where the saturation magnetization of BFO nanoparticles with the mean size of 41 nm was about 0.8 Am$^2$/kg. The decrease in magnetization observed in STC nanoflowers after calcination can be caused by an increase in mean size of crystallites (see table I) or by the crystallization of the amorphous phase. The amorphous phase exhibits higher magnetization that that of crystalline BFO phase due to spin glass behavior [55] and its reduction can lead to a decrease in the total magnetization of the STC nanoflowers.

Temperature variation of the real $\varepsilon$' and imaginary $\varepsilon$'' parts of dielectric permittivity dependence on temperature for ST and SCT samples are shown in Fig. 13.

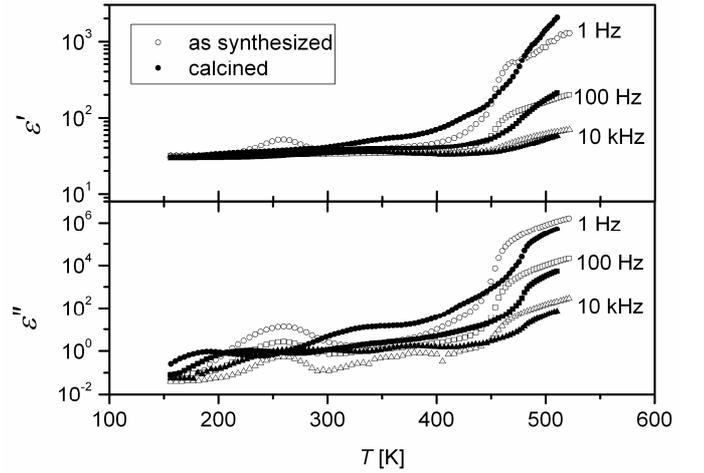

Figure 13. Real $\varepsilon$' and imaginary $\varepsilon$'' part of permittivity for as prepared ST sample (open symbols) and calcined STC sample (solid symbols).

Two main features can be found in this figure – a broad maximum around 200-300 K apparent in $\varepsilon$'' and also in $\varepsilon$' dependences (for low frequency only) and an increase in both real and imaginary parts of the permittivity above about 450 K. The low temperature broad maximum is probably due to water or ice surface condensation [56] or due to the water confined to nanocavities of the samples because of the wet chemistry synthesis. To check this interpretation, STC calcined sample with reduced water content was measured. It turns out that the broad maximum is suppressed in STC calcined sample, which supports our interpretation of the origin of low temperature maximum being due to the water confined to nanovoids. The strong increase in both components of permittivity above 450 K can be interpreted in terms of Polomska transition [57, 58] related to the surface phase transition and anomalous magnon damping [56].

## IV. CONCLUSIONS

In summary, for the first time we report the method of the microwave assisted wet chemistry synthesis of bismuth ferrite multiferroic nanoflowers and the necessary conditions promoting the formation of these nanoobjects. The morphology of nanoflowers is sensitive to KOH content and the time of microwave processing, so that homogeneous powders composed of nanoflowers only are obtained for KOH concentration of 6 M and 20 min synthesis. Different KOH contents lead to formation of other BFO structures, whereas the long-time synthesis leads to powders containing large nanoflowers but also irregular BFO agglomerates.

At early stages of the growth process the central part of nanoflowers or bottom of calyx composed of only a few petals is formed. The next petals are attached successively to this structure as the reaction continues. The finest petals with thickness of about 40 nm are composed of single layer of BFO nanocrystals and also some amount of the amorphous phase.

The size of the nanocrystals forming the petals is smaller that the period of spin cycloid in bismuth ferrite which leads to enhanced magnetization of nanoflowers when compared to that of bulk BFO materials. Dielectric properties of the nanoflowers are dominated by a broad maximum at 200 K ÷ 300 K and an increase in both real and imaginary component of the dielectric permittivity above 450 K. The low temperature maximum can be interpreted in terms of water confined to nanovoids and the increase in permittivity above 450 K is related to Polomska transition.

## V. ACKNOWLEDGEEMENTS

This project has been supported by National Science Centre (project No. N N507 229040) and partially by COST Action MP0904.